\documentclass[
twocolumn,
pre,
reprint
]{revtex4}
\usepackage{amsmath,amssymb}
\usepackage{epsfig}
\usepackage{graphicx}
\usepackage{amsmath}
\usepackage{slashed}
\usepackage{amsfonts}
\usepackage{epstopdf}
\usepackage{ulem}
\usepackage{multirow,tabularx}

\usepackage{xcolor}

\newcommand{\be}{\begin{equation}}
\newcommand{\ee}{\end{equation}}
\newcommand{\bear}{\begin{eqnarray}}
\newcommand{\eear}{\end{eqnarray}}
\newcommand{\ba}{\begin{array}}
\newcommand{\ea}{\end{array}}

\newcommand{\zbar}{\bar{z}}
\newcommand{\wbar}{\bar{w}}
\newcommand{\tr}{\mathrm{Tr}}

\newcommand{\ket}[1]{\left| #1 \right>}
\newcommand{\bra}[1]{\left< #1 \right|}
\newcommand{\braket}[2]{\left< #1 | #2 \right>}

\def\be{\begin{eqnarray}}
\def\ee{\end{eqnarray}}

\def\roughly#1{\mathrel{\raise.3ex\hbox{$#1$\kern-.75em%
\lower1ex\hbox{$\sim$}}}}

\def\d{\mathbb{D}}
\def\Q{\mathbb{Q}}
\def\R{\mathbb{R}}
\def\P{\mathbb{P}}
\def\G{\mathbb{G}}
\def\B{\mathbb{B}}
\def\X{\mathbb{X}}
\def\dQ{\d_{\Q}}
\def\be{\begin{eqnarray}}
\def\ee{\end{eqnarray}}

\begin{document}

\title{Eikonal formulation of large dynamical random matrix models }

\author{ Jacek Grela$^1$, Maciej A. Nowak$^{1,2}$  and Wojciech Tarnowski$^1$}

\email{jacek.grela@uj.edu.pl}
\email{maciej.a.nowak@uj.edu.pl} 
\email{wojciech.tarnowski@doctoral.uj.edu.pl}
\affiliation{
$^1$Institute of Theoretical Physics, Jagiellonian University, 30-348 Cracow, Poland\\
$^2$Mark Kac  Complex Systems Research Center, Jagiellonian University, 30-348 Cracow, Poland
}


\date{\today}
\begin{abstract}

The standard approach to dynamical random matrix models relies on the description of trajectories of eigenvalues.  Using the analogy from optics, based on the duality between the  Fermat  principle (rays) and the Huygens principle (wavefronts), we formulate the Hamilton-Jacobi dynamics for large random matrix models.  The resulting equations describe a broad class of random matrix models in a unified way, including normal (Hermitian or unitary) as well as strictly non-normal dynamics. This formalism applied to Brownian bridge dynamics allows one for calculations of the asymptotics of the Harish-Chandra-Itzykson-Zuber integrals.

\end{abstract}



\maketitle

\setcounter{footnote}{0}

In this work we study matrices undergoing additive or multiplicative random dynamics \cite{DYNBOOK}. Besides their purely theoretical appeal, such models have proven useful in various problems spanning from quantum mechanics to machine learning. Typical examples are disordered mesoscopic wires where the dynamical time $t$ is identified with wires' length. In quantum systems with broken time reversal symmetry $t$ is in turn related to the external magnetic field \cite{EXAMPLE1}. In two-dimensional QCD, the time parameter corresponds to the area of the loop configuration \cite{GOPAKUMAR}, whereas in quantum gravity it is interpreted as the size of the string \cite{EXAMPLE3}. Remarkably, the variable in which the dynamics take place is seldom related to the physical time. Other applications share such exotic interpretations as the time-like variable is the depth of the neural network \cite{EXAMPLE2} or the strength of noise in signal-plus-noise statistical models \cite{CDAT}.


\begin{figure}
    \centering
    \includegraphics[scale=0.29]{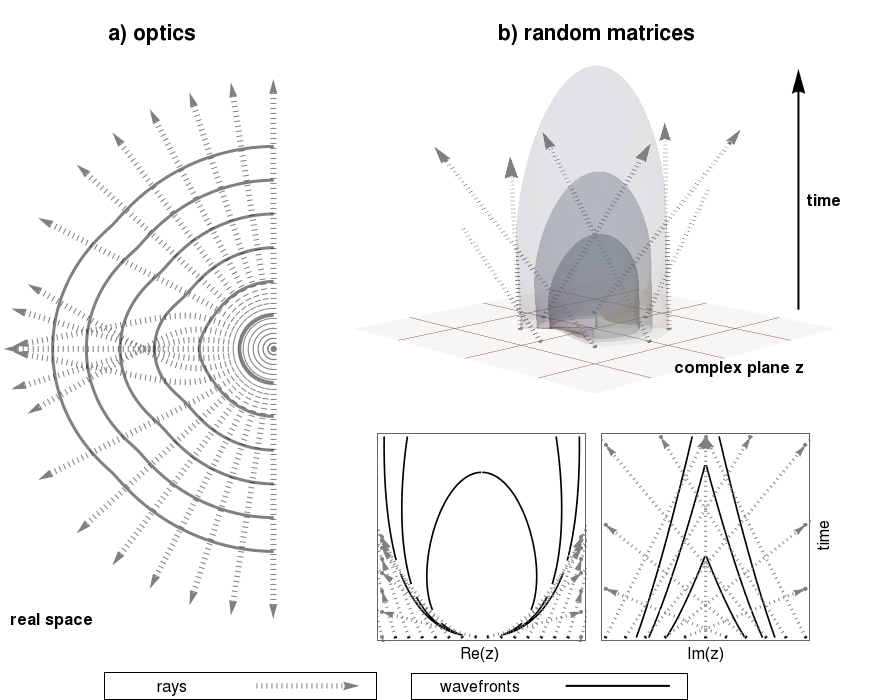}
    \caption{Construction of the duality between wavefronts (black solid lines) and rays (dotted gray arrows) in a) optics and b) dynamical random matrices is highly analogous. Optical wavefronts in real space trace light-ray propagation, while matrix wavefronts $\Phi(z,t) = \text{const}$ trace complex plane propagation of characteristics. In the latter, insets show highly anisotropic wavefront/ray evolution when projected along real and imaginary axes.The presented random matrix dynamics is given by the Gaussian diffusion with zero initial condition giving rise to GUE. Details are given in App. \ref{guedyn}.}
    \label{fig1}
\end{figure}

The aim of this work is to describe the classical mechanics perspective on dynamical matrices as the third natural interpretation besides the hydrodynamical and optical studied previously. Hydrodynamical picture is based on the standard approach to dynamical random matrix models due to Dyson and relies on tracing the trajectories of individual eigenvalues via stochastic differential equations of the Langevin type or by the corresponding Smoluchowski-Fokker-Planck (SFP) equations for joint eigenvalue probability distribution functions. In the limit of large dimension of matrices, $N \to \infty$, the dynamics of random matrices simplifies considerably and attains hydrodynamical description with parameter $1/N$ being the viscosity of the flow of the eigenvalue fluid~\cite{US,NEUBERGER}. 

\begin{table*}[ht]
\begin{ruledtabular}
\caption{\label{tab0}Relations between optics and random matrices}
\begin{tabular}{cc}
    fold/edge/Airy singularity & cusp/Pearcey singularity \\
    \includegraphics[scale=.7]{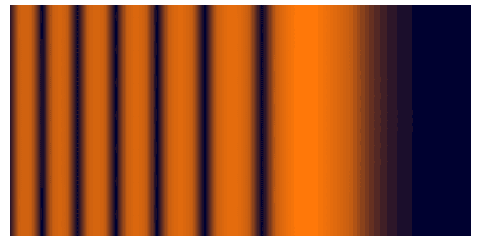} & \includegraphics[scale=.7]{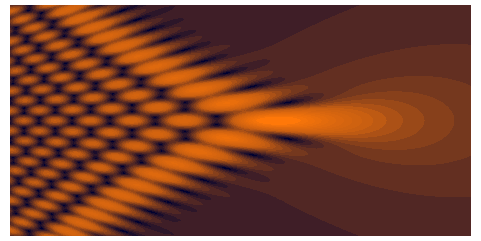} \\
    \hline
    optics & random matrices  \\
    \hline
    wavelength $\lambda$ & inverse of matrix size $1/N$ \\
    geometric optics $\lambda = 0$ & global/macroscopic scaling \\
    wave optics $\lambda \to 0$ & local/microscopic scaling \\
    light intensity $E(x,y) \sim \frac{1}{\lambda^\mu} \Gamma \left (\frac{x}{\lambda^{\sigma_x}}, \frac{y}{\lambda^{\sigma_y}} \right)$ & characteristic determinant $D(z,t)$\\
    fold singularity $\mu = 1/6, \sigma_x = 2/3, \sigma_y = 0$ & edge singularity $D(z,t) \sim N^{1/6} \text{Ai} \left ( \left ( \frac{z}{\sqrt{t}} - 2 \right ) N^{2/3} \right )$\\
    cusp singularity $\mu = 1/4, \sigma_x = 1/2, \sigma_y = 3/4$ & Pearcey singularity $D(z,t) \sim N^{1/4} \text{P}\left ( \frac{t-1}{2} N^{1/2}, z N^{3/4} \right )$
\end{tabular}
\end{ruledtabular}
\end{table*}

Besides the hydrodynamics, an optical analogy in dynamical matrices was likewise established. In the simplest case of Gaussian Unitary Ensemble (GUE), the resolvent evolves according to the complex Burgers equation. It can be easily solved by the method of complex characteristics, in analogy to real characteristics method applied to the Euler equation.  This immediately brings connotations with geometric optics, where rays of light play the role of characteristics. Moreover, fold and cusp diffraction catastrophes in optics~\cite{BERRY} seem to have their counterparts in random matrix models, in terms of Airy~\cite{NOWBLA} and Pearcey~\cite{US} microscopic universalities. We summarize these findings in Tab. \ref{tab0}. 

As a motivation to present work, we utilize the longstanding ray/wavefront duality in optics between Fermat and Huygens (also known as the geometric vs. wave optics) which essentially mirrors the formalisms of Hamilton and Hamilton-Jacobi \cite{ARNOLD}. In the context of dynamical matrices, our aim is to bring into forefront the mechanics perspective with special emphasis on the Hamilton-Jacobi formalism.

Before we present the full formalism for the HJ equation in Random Matrix Theory, let us explain the main concept on the basis of a matricial additive Brownian walk. We consider the process $Y_t=Y_{t-1} +X_t$, where $X_t$ are independent large ($N \rightarrow \infty$) $N$ by $N$ matrices drawn from the GUE. 
When interested only in the average spectral density, one studies the evolution of the averaged resolvent $G(z,t)= \left < \frac{1}{N}{\rm Tr }\frac{1}{z-Y_t} \right >$, with the large $N$ limit taken implicitly. The averaging $\left < \cdots\right >$ is taken with respect to the random process $Y_t$. In the above-mentioned limit,  the resulting differential equation is the complex inviscid Burgers equation
$\partial_t G +G\partial_zG=0$~\cite{BURGERS}. Using the method of complex characteristics (shown as gray dotted arrows in Fig. \ref{fig1}b), the solution is given implicitly by the Pastur formula~\cite{PASTUR} $G=G_0(z-tG)$, where $G_0$ is the initial resolvent. For a trivial initial condition $X_0=0$, $G_0(z)=1/z$ and the Pastur formula reduces to a quadratic equation for which one of the solutions $G_-(z,t)=\frac{1}{2t}(z-\sqrt{z^2-4t})$ results in the eigenvalue density given by the Wigner semicircle law. In this approach, we omit the wavefronts altogether.


However, following Huygens, the picture will be complete only when we recast the problem in the HJ form. Then the role of the principal Hamilton function is played by a potential-like function of the form $ \left < \frac{1}{N} {\rm Tr}\ln (z-Y_t)(\bar{z}-Y_t^{\dagger} ) \right >$ with large $N$ limit taken implicitly. Its equipotential surfaces are precisely the omitted wavefronts, shown as black solid lines/surfaces in Fig. \ref{fig1}b.

Since $Y_t$ is Hermitian, the proposed principal Hamilton function is decomposed as a sum of holomorphic $\phi(z,t)=  \left <  \frac{1}{N} {\rm Tr} \ln (z-Y_t) \right >$  and its (trivial) anti-holomorphic copy $\bar{\phi}=\phi(\bar{z},t)$, which we omit in what follows. Moreover, the function $\phi$ is basically a logarithm of the characteristic determinant, since 
$\left < {\rm Tr}  \ln (z-Y_t) \right >= \left <\ln \det (z-Y_t) \right > = \ln \left <\det(z-Y_t) \right >$, where the last equality holds {\it only} in the $N \to \infty$ limit. The HJ equation for the principal Hamilton function (modulo its trivial, decoupled anti-holomorphic copy) reads
\begin{align*}
 \partial_t \phi  +H(p=\partial_z\phi, z, t)=0,
\end{align*}
where $H(p,z,t) = p^2/2$ is the Hamiltonian. The role of the canonical coordinate $q$ is played by a complex variable $z$, while the role of the canonical momentum $p$ is the derivative of the principal Hamilton function wrt. coordinate $z$, i.e. $p=\partial_z\phi$. Note that the momentum $p$ is, by definition, the resolvent $G$! Surprisingly from the random matrix point of view, the HJ formalism treats the canonical pair $(q,p) \leftrightarrow(z,G)$  as completely {\it independent}. As we will see in the next section, GUE is the random matrix  analog of free, 1-dimensional particle in classical mechanics (see Tab. \ref{tab1}). 

Using the formalism of classical mechanics, we write down the pair of Hamilton equations $\dot{z}=\frac{\partial H}{\partial p}=p$, $\dot{p}=-\frac{\partial H}{\partial z}=0$, which, together with the initial conditions $z(0)=z_0$ and $p(0)=p_0$, lead to the solutions $p(t)=p_0$ and $z(t)=p_0t+z_0$. In accordance with previous ray-centered appraoch, the latter equation give the characteristics. If initial conditions are represented by a set of $N$ points $x_i$ corresponding to the eigenvalues of $X_0$, then $p_0=\partial_z  \phi(z,t=0)|_{z=z_0}=\frac{1}{N} \sum \frac{1}{z_0-x_i}$ and eliminating $z_0$ from equations of motion reproduces the Pastur formula $p=p_0(z-pt)$. Alternatively, one can differentiate the HJ equation with respect to $z$, again recovering the inviscid Burgers equation $\partial_t p+p\partial_z p=0$.

The main result of present work is an extension of the above duality to broader class of dynamical random matrix models, not necessarily Hermitian or Gaussian. 



\begin{table*}[ht]
\caption{\label{tab1a} Hamiltonians for HJ equation \eqref{eq:HJHerm} describing Hermitian additive matrix dynamics. Besides the bridge scenario discussed in Sec. \ref{sec:bridge} where phase space comprises of two complex pairs $(z,p), (\alpha,p_\alpha)$ and the Hamiltonian is non-stationary, all remaining examples are defined for a single complex pair $(z,p)$ and do not depend explicitly on time $t$.}
\begin{ruledtabular}
\begin{tabular}{lll}
 dynamics type & Hamiltonian &  details \\
\hline
 GUE & $p^2/2$ & \\ 
 R-based & $\int_0^p R_X(z) dz$ &  a general R-transform \\  
 Ornstein-Uhlenbeck & $\frac{1}{2}p^2 +a(1-zp)$ & $a$ is the drift parameter \\
 bridge & $ \frac{1}{2} p^2 + \frac{1}{1 - t} \left [ 1 - zp - (\alpha-1)p_\alpha \right ]$ & details in Sec. \ref{sec:bridge}\\
 Wishart & $(1-r)p+rzp^2$ & $r$ is the rectangularity parameter \cite{WISHART} \\
 Jacobi & $\lambda\theta z(1-z)p^2+p[\theta(1-\lambda) -(1-2\lambda\theta)z]$ & $\theta, \lambda$ are defined in~\cite{JACOBI} \\
\end{tabular}
\end{ruledtabular}
\end{table*}

\section{Additive matrix dynamics}
We define a general additive matrix process $Y_t$ by:
\begin{equation}
\label{defherm}
R_{Y_t}(z,t)=R_{X_0}(z)+tR_{X}(z)
\end{equation}
with R-transform $R_X$ for the increment matrix and an initial matrix $X_0$. In the above, large matrix size limit was taken implicitly. Although it is advantageous to introduce this quite involved definition of matrix process, one can think of it in simpler terms by considering the finite dimensional case. For Gaussian increments, formulation \eqref{defherm} is equivalent to a limit of additive process $Y_t=X_0+X_1+X_2+X_3+\ldots +X_n$ with independent finite-dimensional increments $X_i$ each with variance $\delta t$ where the limit $n \to \infty, \delta t \to 0$ is taken with $n\delta t = t$ fixed. Beyond Gaussianity, similar definition is also possible although it is slightly more involved as the increments contain additional random projections on top of the $X_i$'s \cite{BELNICA}. 

\subsection{Hermitian Hamilton-Jacobi equation}
\label{sec:herm}
We first look at case when the increment matrix $X$ is Hermitian. Unless otherwise stated, we work in the limit $N\to \infty$. To derive the Hamilton-Jacobi equation, we introduce several well-known results of free probability applicable to Hermitian matrices. 

\textit{Basics of free probability.}
The one-point spectral density $\rho(\lambda)=\left<\frac{1}{N} \sum_{i=1}^{N}\delta(\lambda-\lambda_i)\right>$ is uniquely given by its Stjelties (Cauchy) transform, also known in physics literature as Green's function $G(z)=\int \rho(\lambda) (z-\lambda)^{-1}d\lambda=\left<\frac{1}{N}\tr (z-X)^{-1}\right>$, which encodes all its moments. One recovers the spectral density by the Sochocki-Plemelj formula:
\begin{equation*}
\rho(\lambda)=-\frac{1}{\pi} \lim_{\epsilon\to 0^{+}} \text{Im}\, G(\lambda+i\epsilon).
\end{equation*}
Free probability offers several operational tools to deal with spectra of asymptotically large matrices. In particular, with the use of freeness (a counterpart of independence in non-commuting random variables), one is able to find an eigenvalue density of a sum of two matrices by knowing their separate densities. 

To this end, one introduces a functional inverse of the Green's function, called Blue's function, satisfying
\begin{equation}
B(G(z))=z, \qquad G(B(z))=z, \label{GreenBlue}
\end{equation}
which is an intermediate step to calculate the $R$-transform $R(z)=B(z)-1/z$. Remarkably, $R$-transform is additive for two mutually free random variables $X$ and $Y$, that is~\cite{Voic}
\begin{equation*}
R_{X+Y}(z)=R_X(z)+R_Y(z).
\end{equation*}
This additive property explains and enables the definition of matrix process in terms of R-transform given in eq. \eqref{defherm}.

\subsubsection{Derivation of HJ equation}
Firstly, we add $1/z$ to both sides of the equation \eqref{defherm} resulting in  $B(z,t)=B_{X_0}(z)+tR_{X}(z)$ which we then differentiate with respect to $t$ and obtain
\begin{align}
\partial_t B(z,t)=R_X(z).
\label{eq0}
\end{align}
Since the matrix $Y_t$ evolves, both Green's and Blue's functions depend on time, but the relation \eqref{GreenBlue} is satisfied at \textit{any} time:
\begin{equation}
B_{Y_t}(G_{Y_t}(z,t),t)=z. \label{blueandgreen}
\end{equation}
From now on we skip the subscripts $B_{Y_t} \to B, G_{Y_t} \to G$. Differentiating the above definition with respect to $t$, we get
\begin{equation}
\left.\partial_t B(z,t)\right|_{z=G}+\left.\frac{\partial B(z,t)}{\partial z}\right|_{z=G}  \partial_t G(z,t)=0. \label{dynamics1}
\end{equation}
On the other hand, we differentiate eq. \eqref{blueandgreen} with respect to variable $z$ $\left.\frac{\partial B(z,t)}{\partial z}\right|_{z=G}\frac{\partial G(z,t)}{\partial z}=1$ and substitute it to formula \eqref{dynamics1} to obtain $\left.\partial_t B(z,t)\right|_{z=G}+\left(\frac{\partial G}{\partial z}\right)^{-1}   \partial_t G(z,t)=0$.
Knowing  the time derivative of the Blue's function \eqref{eq0}, we finally arrive at
\begin{equation}
\label{Voic}
\partial_t G(z,t)+R_X(G) \partial_z G(z,t)=0.
\end{equation}
which is the Voiculescu equation.

\textit{Formal solution of Voiculescu equation.}
We proceed to formally solve equation \eqref{Voic} by the method of characteristics. The result are two equations:
\begin{align*}
\partial_t z & = R_X(G), \\
\partial_t G & = 0, 
\end{align*}
which we interpret as Hamilton equations $\dot{q} = \partial_p H, \dot{p} = - \partial_q H$ where pair $(z,G)$ becomes the coordinate-momentum pair $(q,p)$. Then the Hamiltonian is specified by $\partial_G H = R_X(G), \partial_z H = 0$ which gives
\be
\label{mainherm}
H(G,z) = \int^G_0 dz R_X(z).
\ee
Lower integration limit is a convention introduced to fix a constant term in the Hamiltonian. Knowing the Hamiltonian, we write down the corresponding Hamilton-Jacobi equation for Hamilton's principal function $S$ as $\partial_t S = - H\left (\partial_z S, z \right )$ and take the derivative wrt. $z$ which results in the Voiculescu equation \eqref{Voic} with identification $\partial_z S = G$. Hence, we idengtify the principal function $S$ with the electrostatic potential
\begin{align*}
    \phi(z,t) = \frac{1}{N} \left < \text{Tr} \ln (z-Y_t) \right > 
\end{align*}
as $\partial_z \phi = G$. As a result, the sought Hamilton-Jacobi equation for the electrostatic potential reads
\begin{align}
\label{eq:HJHerm}
    \partial_t \phi + H\left (p = \partial_z \phi, z \right ) = 0 .
\end{align}

\textit{Comments.}
Although Hamiltonians expressed in the most general form via the R-transform \eqref{mainherm} are functions of momenta only, the HJ equation \eqref{eq:HJHerm} holds beyond such cases. Perhaps the simplest instance is the Hamiltonian
 \be H_{\text{OU}}=\frac{1}{2}p^2 +a(1-zp),
 \label{HOU}
 \ee
 where the coupling between coordinate $z$ and momentum $p$ reproduces the Ornstein-Uhlenbeck process with a drift proportional to $a$ \cite{OU}. 
 
 In Tab. \ref{tab1a} we summarize examples of Hamiltonians considered throughout this work, supplemented with Jacobi and Wishart processes.
 
\begin{table*}[ht]
\caption{\label{tab2a} Hamiltonians for HJ equation \eqref{eq:NHHJ}  describing non-Hermitian additive matrix dynamics. }
\begin{ruledtabular}
\begin{tabular}{ c c c }

 dynamics type & Hamiltonian &  details \\
\hline
 Ginibre & $-|p_w|^2$ & \\ 
 R-based & $\int_{0}^{{ \P}}  \rm Tr \left [ \R^X\left (\Q \right ) d\Q \right ]$ & a general non-Hermitian $\R$-transform \\  
 elliptic & $\frac{\tau}{2}(p_z^2+p_{\bar{z}}^2)-|p_w|^2,$ & $\tau$ is the interpolation parameter \\
 R-diagonal & $\int_0^{-|p_w|^2} A(x) dx$ & $A$ is a cumulant generating function\\
\end{tabular}
\end{ruledtabular}
\end{table*}
 
\subsection{Non-Hermitian Hamilton-Jacobi equation}  
We now continue to outline the HJ formalism in the case when the matrix dynamics $Y_t$ is still additive but lacks symmetry constraints. We start off with free probability terms and definitions applicable to this scenario.

\textit{Non-Hermitian free probability.}
Eigenvalues of diagonalizable (not necessarily normal) random matrices form a subset of the complex plane. In order to work with such objects we use the following representation of the Dirac delta~\cite{Crisanti,Janik,Feinberg,Chalker}: 
\begin{equation*}
\delta^{(2)}(z)=\frac{1}{\pi}\lim_{\epsilon\to 0}\frac{\epsilon^2}{(z\bar{z}+\epsilon^2)^2}.
\end{equation*}
In the spirit of the electrostatic analogy, one introduces the potential:
\begin{equation}
\Phi(z,w,t)=\left<\frac{1}{N} \text{Tr} \ln \left [(z-Y_t)(\bar{z}-Y_t^{\dagger})+|w|^2\right ] \right>. \label{potential}
\end{equation}
The limiting spectral density can be recovered from the Poisson law:
\begin{equation*}
\rho(z)=\lim_{w\to 0}\frac{1}{\pi}\partial_{z\bar{z}}\Phi(z,w,t).
\end{equation*}
Using known identity $\text{Tr} \ln = \ln \det$, the determinant in \eqref{potential} can be rewritten in block form:
\begin{equation*}
\Phi(z,w,t)=\left<\frac{1}{N}\ln\det(\Q-\X)\right>, 
\end{equation*}
where
\begin{equation*}
\Q=\left(\begin{array}{cc}z & -\wbar \\ w & \zbar
\end{array}\right), \qquad \X=\left(\begin{array}{cc}
X & 0 \\ 0 & X^{\dagger}
\end{array}\right).
\end{equation*}
$\Q$ is a $2\times 2$ matrix representation of the real quaternion. In direct analogy to Hermitian matrices, one constructs the Green's function of a quaternion argument which is now a $2\times 2$ matrix:
\begin{equation}
{\G}({\Q}) = \dQ \Phi = \left(\begin{array}{cc}
\partial_z \Phi &  \partial_w \Phi \\ -\partial_{\wbar} \Phi & \partial_{\zbar} \Phi
\end{array}\right). \label{eq:GreenQuaternion}
\end{equation}
with a quaternionic derivative $(\dQ)_{ij} \equiv \frac{\partial}{\partial \Q_{ji}}$ ($i,j=1,2$).
In direct analogy, inverse of quaternionic Green's function is the non-Hermitian analog of the Blue's function ${\B (\G(\Q))=\Q=\G(\B(\Q))}$. This directly leads to the quaternionic $\R$-transform ${\R(\Q)= \B(\Q)-\Q}^{-1}$. As previously, it is additive under addition of free non-Hermitian matrices~\cite{JaroszNowak}:
\begin{equation}
\label{nhadd}
{\R}_{X+Y} ({\Q} )= {\R}_{X}({\Q})+{\R}_{Y}({\Q}).
\end{equation}

A generalized resolvent $\G$ was proposed previously to solve non-Hermitian problems in the past~\cite{USQUAT,QUATBLUE,FEINBERZEE,CHALKERWANG} although without any link to the underlying Hamilton dynamics. 

\subsubsection{Deriving the HJ equation}

We use indices $\alpha,\beta=1,2$ to specify matrix elements. Non-Hermitian additive dynamics is defined by eq. \eqref{defherm} with straightforward substitutions $R \to \R, z \to \Q$ motivated by additive property \eqref{nhadd}. We add $\Q^{-1}$ to resulting equation so that
\begin{equation*}
\B_{\alpha\beta}(\Q,t)=\B^{0}_{\alpha\beta}(\Q)+t \R^{X}_{\alpha\beta}(\Q),
\end{equation*}
where $\B = \B^{Y_t}$ is the Blue's function for the matrix $Y_t$ while $\R^X$ is the $\R$-transform of the increment matrix $X$ with standard variance. We again calculate the time derivative 
\begin{equation}
\partial_t \B_{\alpha\beta}(\Q,t)=\R_{\alpha\beta}^{X}(\Q).\label{Rtransf}
\end{equation} 
It is convenient to treat quaternionic objects not as $2\times 2$ matrices but as column vectors with 4 components in, let us say, lexicographic order: $\Q_{\alpha}=(\Q_{11},\Q_{12},\Q_{21},\Q_{22})^T$. Now $\alpha=1,2,3,4$. Such vector representation makes derivation less convoluted. The quaternionic Blue's $\B_{\alpha}(\Q,t)$ and Green's $\G_{\alpha}(\Q,t)$ functions are, by definition, related as:
\begin{equation*}
\B_{\alpha}(\G(\Q,t),t)=\Q_{\alpha}.
\end{equation*} 
As previously, we differentiate above definition wrt. time to get
\begin{equation}
\left.\partial_t \B_{\alpha}(\Q,t)\right|_{\Q=\G} + \sum_{\beta=1}^{4} \left.\frac{\partial \B_{\alpha}(\Q,t)}{\partial \Q_{\beta}}\right|_{\Q=\G}\frac{\partial \G_{\beta}(\Q,t)}{\partial t}=0,\label{timederivative}
\end{equation}
and with respect to the quaternionic element $\Q_\beta$:
\begin{equation*}
\sum_{\beta=1}^{4} \left.\frac{\partial \B_{\alpha}(\Q,t)}{\partial \Q_{\beta}}\right|_{\Q=\G}\frac{\partial \G_{\beta}(\Q,t)}{\partial \Q_{\gamma}}=\delta_{\alpha\gamma}.
\end{equation*}
We see that the above matrices of derivatives are mutual inverses. Multiplying \eqref{timederivative} on the left by $\partial \G_{\gamma}/\partial \Q_{\alpha}$, summing over repeated indices and substituting the expression \eqref{Rtransf} we are led to
\begin{equation*}
\partial_t \G_{\alpha} + \sum_{\beta=1}^{4} \R^{X}_{\beta}(\G)\frac{\partial \G_{\alpha}}{\partial \Q_{\beta}}=0.
\end{equation*}
Finally, we restore the quaternionic structure to arrive at generalized Voiculescu-type equation:
\begin{equation}
\partial_t \G_{\alpha\beta} + \sum_{\mu,\nu =1}^2 \R^{X}_{\mu\nu}(\G)\frac{\partial \G_{\alpha\beta}}{\partial \Q_{\mu\nu}}=0 \label{NonhermitianDynamics}.
\end{equation}
which is a direct generalization of Voiculescu equation \eqref{Voic}.

\textit{Formal solution.} As previously, obtained equation is amenable to solving by the method of characteristics which result in first order ODEs:
\begin{align*}
    \dot{\Q}_{\mu\nu} & = \R^X_{\mu\nu} (\G), \\
    \dot{\G}_{\mu\nu} & = 0,
\end{align*}
where $\mu,\nu = 1,2$. Above equations are again in Hamilton form, where the pair $(\Q,\G)$ is identified with a set of coordinate-momentum pairs $(\Q,\P^T)$. Transposition is indispensable to align the Green's function \eqref{eq:GreenQuaternion} as a derivative of potential wrt. quaternion $\Q$. As a consequence, first equation reads $\dot{\Q}_{\mu\nu} = \R^X_{\nu\mu} (\P)$ following from $\R (\P^T) = \R (\P)^T$. Hamiltonian $H(\P,\Q)$ is found from equations $\partial_{\P_{\nu\mu}} H = \R^X_{\mu\nu} (\P)$ and $\partial_{\Q_{\mu\nu}} H = 0$. We integrate out each one separately so the result is a sum of integrals:

\begin{align}
    H(\P, \Q) & = \int_0^{\P} d\Q_{11} \R^X_{11}(\Q) + \int_0^{\P} d\Q_{12} \R^X_{21}(\Q) + \nonumber \\
    & + \int_0^{\P} d\Q_{21} \R^X_{12}(\Q) + \int_0^{\P} d\Q_{22} \R^X_{22}(\Q), \label{NHH}
\end{align}
where the lower limit is again introduced to fix an arbitrary additive constant in the Hamiltonian.
Instead of multiple terms present in \eqref{NHH}, in what follows we introduce a succinct notation
\begin{eqnarray}
H(\P,\Q) = \int_{0}^{{ \P}}  \rm Tr \left [ \R^X\left (\Q \right ) d\Q \right ].
\label{main}
\end{eqnarray}
Newfound Hamiltonian admits the following Hamilton-Jacobi equation for Hamilton's principal function $S$
\begin{align*}
    \partial_t S + H \left ( \P = (\dQ S)^T, \Q \right ) = 0.
\end{align*} 
Lastly, we identify principcal function with a known matrix object. To this end, we take the derivative $\dQ$ so that $(\dQ)_{kl} H \left ( \P, \Q \right ) = \sum_{ij} \R^X(\P)_{ij} (\dQ)_{kl} \P_{ji}$ and
\begin{align*}
\partial_t (\dQ)_{kl} S + \sum_{ij} \R^X (\dQ S^T)_{ij} (\dQ)_{ij} (\dQ)_{kl} S = 0.
\end{align*}
We use again $\R(\P^T) = \R(\P)^T$ so that
\begin{align*}
\partial_t (\dQ)_{kl} S + \sum_{ij} \R^X (\dQ S)_{ji} (\dQ)_{ij} (\dQ)_{kl} S = 0.
\end{align*}
Since $(\dQ)_{ij} = \partial_{\Q_{ji}}$, we recreate the Voiculescu-type equation \eqref{NonhermitianDynamics} when the principal function $S$ is identified with electrostatic potential $\Phi$ \eqref{potential} so that $\dQ S = \G$. 

As a result, we have derived the main result of this paper, a Non-Hermitian Hamilton-Jacobi equation:
\begin{align}
\label{eq:NHHJ}
    \partial_t \Phi + H \left ( \P = (\dQ \Phi)^T, \Q \right ) = 0
\end{align} 
matches eq. \eqref{NonhermitianDynamics}. Hamilton equations are readily solved as $\P = \P_0 (\Q-t\R(\P)^T)$ with initial condition $\P_0 = \dQ \Phi^T$. Coordinates $\Q,\P$ comprise a set of action-angle variables casting the problem as fully integrable and stable wrt. small perturbations according to the seminal KAM theorem \cite{ARNOLD}.

\subsubsection{Examples}
\label{sec:nhexamples}
In this section we consider few examples of descriptions of non-Hermitian additive matrix dynamics in terms of Hamilton-Jacobi equations. We provide a summary Tab. \ref{tab2a} where we present examples of discussed Hamiltonians for non-Hermitian HJ equation \eqref{eq:NHHJ}.

\paragraph{Reduction to Hermitian case.} 
Non-Hermitian formalism presented in this section is not disjoint from the Hermitian dynamics considered in Sec. \ref{sec:herm}. Now we show that in fact, it is contained within Non-Hermitian framework at least in part. In the case of Hermitian matrices, the quaternionic embedding is redundant and one can set $w$ to zero from the very beginning, projecting the quaternion to a complex number. In this way, both the potential $\Phi \to \phi + \bar{\phi}$ and the quaternion $\R = \text{diag}(R_X(z),\overline{R_X(z)})$ decouple into holomorphic and anti-holomorphic copy. The Hamiltonian \eqref{main} likewise decouples and reads 
\begin{equation*}
H = \int_0^{p} R_X(z)dz +\int _0^{\overline{p}} \overline{R_X(z)} d\bar{z},    
\end{equation*}
so the dynamics of each part separately is equivalent; the holomorphic part of the Hamiltonian  recreates exactly the eq. \eqref{mainherm} found in the Hermitian scenario.
  
\paragraph{Non-normal increment matrix $X$.} 
The crucial difference between Hermitian and non-Hermitian models comes from the fact, that the  holomorphic/anti-holomorphic separability breaks down, since the support of the spectra represents the non-holomorphic region. This was known in the literature~\cite{CRISANTI}, and the variable $|w|^2\equiv \epsilon$ was kept non-zero before the large $N$ limit was taken. In such a case, the spectral density follows from the 2D Gauss law $\rho =\frac{1}{\pi}  \partial_{\bar{z}} g $, where $g=\partial_z\Phi$ plays the role of the electric field. Considering $\epsilon$ only as an infinitesimal regularizer is too reductive, as it is responsible for the crucial dynamics of eigenvectors,  which, contrary to the Hermitian case, do not decouple from the eigenvalues during the evolution. This is perhaps best visible when we diagonalize $Y_t$ in terms  of left and right eigenvectors $Y_t=\sum_i  \ket{R_i} \lambda_i \bra{L_i} = R\Lambda L^\dagger$.  Then the potential  $\Phi$ reads explicitly 
\begin{eqnarray*}
\Phi(z,w,t) = \frac{1}{N}
\left<
\rm \ln \det
\left(
\begin{array}{cc}
z-{\Lambda} &   -\bar{w} L^\dagger L \\
w R^\dagger R & \bar{z} -{{\Lambda}^{\dagger}}
\end{array}
\right)
\right>.
\end{eqnarray*}
Since the $N$ by $N$ blocks in the determinant do not commute, eigenvalues are correlated with eigenvectors.
In the large $N$ limit 
the off-diagonal momenta in $\G$ are responsible for the diagonal part of the Chalker-Mehlig correlator~\cite{CHALKERMEHLIG,NN,SPEICHER}
\begin{equation*}
O(z,t)= \frac{1}{N^2} \left<  \sum_i O_{ii} \delta^{(2)}(z-\lambda_i) \right> =
 - \frac{1}{\pi} |p_w|^2_{w=0}, \nonumber
\end{equation*}
where $O_{ii}$  is the diagonal part of the overlap matrix~\cite{CHALKERMEHLIG}  $O_{ij}=\braket{L_i}{L_j}\braket{R_j}{R_i}$ (see also~\cite{BellSt}).  This quantity is also related to the Petermann factor~\cite{BEENAKKER} and the eigenvalue condition number in the stability theory~\cite{WILKINSON}.
One can therefore see that during the evolution  parameters $z$ and $w$ need to be treated {\it on an equal footing}.

It is useful to illustrate this democracy of dynamics of eigenvalues and  eigenvectors in the case of the elliptic ensemble~\cite{GIRKO}, corresponding to the matricial measure $P(X) \sim 
\exp \left[ -\frac{N}{1-\tau^2}\left({\rm Tr }XX^{\dagger}- \frac{\tau}{2}{\rm Tr} (X^2+(X^{\dagger})^2)\right) \right].$
Parameter $\tau$ allows for continuous interpolation between GUE ($\tau=1$) and the Ginibre ensemble ($\tau=0$).
The generalized $\R$-transform for the elliptic ensemble reads ~\cite{NN,SPEICHER} 
\begin{equation*}
\R^X(\Q)=\left(
\begin{array}{cc} \tau z & -\bar{w} \\
w & \tau \bar{z}
\end{array}
\right).
\end{equation*}
The application of the HJ formula \eqref{main} leads to 
\be
H_{\text{elliptic}}  & = & \int^\P_0 (\tau zdz +\tau \bar{z} d\bar{z} -wd\bar{w} -\bar{w} dw) \nonumber \\ 
& = & \frac{\tau}{2}(p_z^2+p_{\bar{z}}^2)-|p_w|^2,  
\label{Helliptic}
\ee
 with a pair of momenta $p_z=\G_{11}, \, p_w = \G_{12}$ comprising the quaternionic resolvent $\G$.
 Indeed, setting $\tau=1$ reproduces the GUE case as the "eigenvector part" vanishes in the large $N$ limit. 
Although eigenvector and eigenvalue parts in the Hamiltonian are decoupled, they are coupled by the initial condition.  The presence of the $\tau$ part is actually spoiling the rotational symmetry of the Ginibre ensemble  and reproduces the ellipse, as easily seen from solving the corresponding HJ equations. The signs in front of the "kinetic" terms are also important. In the Hermitian limit $\tau=1$, the positive kinetic term in the Hamiltonian is responsible for the Airy oscillations at the wavefront. When Hermiticity is broken, the term $-|p_w|^2$ shapes the critical behavior at the edge and is the source of smooth decay of Erfc type~\cite{OURPRL,OURNPB}.
 
 
 Along the solution of the HJ equation, $H,\frac{\tau}{2}p_z^2,\frac{\tau}{2} p_{\bar{z}}^2$ and $|p_w|^2$ are constants of motion, since the corresponding Poisson brackets vanish. We stress here the crucial dynamics of eigenvectors, which is a generic feature of non-normal random matrix models as argued recently in \cite{GRELAWARCHOL,DUBACH}. 
 
 In the Ginibre case $\tau=0$, the entire evolution of eigenvalues and eigenvectors is solely driven by the $w$ dynamics, and in this simplest non-normal case, by the Chalker-Mehlig  eigenvector correlator. Explicitly, the HJ equations read $\dot{p}_w = 0$, $\dot{w} = - \overline{p}_w$ and form equations along characteristic lines reproducing recent result~\cite{OURPRL}. 

\paragraph{R-diagonal matrices.}

If we consider a random complex number, its probability distribution function can take in general a complicated form. One can consider a simplified pdfs which are effectively one-dimensional. One of the examples are the isotropic random variables, defined as follows. Any complex number can be written in a polar form $z=re^{i\varphi}$. A complex random variable is said to be isotropic if its pdf depends only on $r$. In such a case the pdf for a phase $\varphi$ is uniform on a unit circle, yet $r$ and $\varphi$ are independent.

In the analogy to isotropic complex random variables one considers a class of non-Hermitian random matrices which we call isotropic.  Any matrix $X$ possesses a polar decomposition $X=HU$, where $H$ is Hermitian positive definite and $U$ is unitary. $U$ plays a role of the 'phase' of a matrix, therefore if $X$ was to be isotropic, $U$ has to be distributed uniformly on $U(N)$ group. Such a probability distribution function exists and is called the Haar measure. Moreover, $U$ and $H$ have to be mutually free. In the literature such matrices belong to the bi-unitary ensembles, because the probability density for their elements is invariant under multiplication by two independent unitary matrices from both sides. Mathematically, $P(X) = P(UXV)$ for $U,V\in U(N)$.

In this case the spectral properties of the isotropic matrix are completely determined by the spectral distribution of the 'squared modulus' $XX^{\dagger}$. The precise relation is given by the Haagerup-Larsen theorem~\cite{HaagLars}. Recently this theorem was extended to describe also the eigenvector correlation function~\cite{HaagLarsVect}.

The only non-vanishing cumulants are of the form $\alpha_k=\frac{1}{N}\tr (XX^{\dagger})^k$. Let us define a generating function for all cumulant of such matrices
\begin{equation*}
A(x):=\sum_{k=1}^{\infty}\alpha_k z^{k-1},
\end{equation*}
which is also known under the name of generating sequence.
The quaternionic $\R$-transform of such matrices assumes a remarkably simple form~\cite{HaagLarsDiagr}
\begin{equation*}
\R^X(\Q)=A(-|w|^2)\left(\begin{array}{cc}
0 & -\wbar \\
w & 0 
\end{array}\right).
\end{equation*}
By direct substitution to \eqref{main} we calculate the R-diagonal Hamiltonian as
\begin{align}
\label{Hbiunit}
H_{\text{R-diag}} = \int_0^{-|p_w|^2} A(x)dx.    
\end{align}

\section{Multiplicative matrix dynamics}

HJ equation shows up also for multiplicative matrix dynamics of the form:
\begin{align*}
 Z_t=\prod_{j=1}^M \exp{\sqrt{\delta t} X_j},
\end{align*}
where $X_j$ are independent random matrices. Continuous version of such random walk is defined in the limit $\sqrt{\delta t} \rightarrow 0$, $M \rightarrow \infty$, $M \delta t  = t$ fixed. In general, symmetries of $X_j$ induce two natural classes of such dynamics:
\begin{enumerate}
    \item $X_j = i H_j$ with $H_j$ Hermitian. $Z_t$ is unitary, its eigenvalues lie on the unit circle,
    \item $X_j$ is non-Hermitian. $Z_t$ has complex eigenvalues.
\end{enumerate}
Below we provide examples for each type of dynamics.

\subsubsection{Hermitian multiplicative dynamics}
Let $H_j$ be a Gaussian Hermitian matrix. Since unitary matrices are normal, eigenvectors and eigenvalues decouple.  As eigenvalues of $P_t$ lie on the unit circle, it is convenient to investigate their phases $\lambda_i(t)=
\exp {i\theta_i (t)}$ and consider a potential which respects the $2\pi$ periodicity of the phase
\begin{align*}
\phi(\theta, t) =\frac{1}{N} \left < \sum_{i=1}^{N} \sum_{k \in \mathbb{Z}} \ln (\theta- \theta_i(t) +2k \pi) \right >.
\end{align*}
In this case, the evolution resembles an additive case, modulo that the principal Hamilton function has to take into account the periodicity of the angular  variable.  The conjugate momentum $\partial_{\theta}\phi \equiv J$, is obtained by noticing the series expansion of the cotangent
\begin{align*}
J(\theta)=\frac{1}{2} \int_{-\pi}^{\pi} \cot \frac{(\theta-\varphi)}{2} \rho(\varphi) d\varphi.
\end{align*}
The Burgers equation reads $ \partial_t J+ J\partial_{\theta}J=0$ ~\cite{CepaLepingle}. Equivalently, $\phi(\theta,t)$ evolves according to the HJ equation with the Hamiltonian $H=\frac{J^2}{2}$. This example, where the unitary evolution is represented by the canonical pair (angle $\theta$, angular momentum $J$) is a free rotator.

The same problem can be formulated in $z=e^{i \theta} $ variable~\cite{JanikWiecz} where the principal Hamilton function is given again by the log of the characteristic determinant, but the resulting Hamiltonian is less trivial and reads
\be
H=-\frac{1}{2}z^2p^2+\frac{1}{2}zp. \label{HamJW}
\ee

\subsubsection{Non-Hermitian multiplicative evolution}

The non-normal evolution is highly nontrivial as eigenvectors enter non-trivially into the evolution process. Like its unitary analogue, such evolution also develops a structural phase transition manifested by the change of topology in the support of complex eigenvalues~\cite{EWA,WETTIG,BIANE}. This topological phase transition does not depend on the type of $X_j$ and appears in both Hermitian (GUE) and non-Hermitian (Ginibre) cases. Although the shape of the boundary was explicitly calculated for the Ginibre case in \cite{EWA,WETTIG}, understanding of the spectral density was beyond the reach of mathematical methods available at that time. Only very recently, explicit spectral formulae were calculated by \cite{KEMP,HALL}, using the formalism of the partial differential equations of the HJ type.
Somehow conservatively, the authors concentrated on the spectral evolution, but their Hamiltonian, when rephrased in our language of $(z,w)$  variables, reads explicitly
\begin{align*}
H=\frac{r}{2}p_{r} \left(1+ \frac{|z|^2-r^2}{2r}p_{r}  -z p  - \bar{z} \bar{p} \right),
\end{align*}
where $r = |w|$ is the radial coordinate and $p_r$ its conjugate momentum. Clearly, the dynamics is driven primarily by the $w$-evolution (eigenvectors), coupled non-trivially to the $z$-evolution (eigenvalues). 

Interestingly, the HJ equation can be applied to the singular value problem of this non-Hermitian evolution with $X_j$ drawn from Ginibre ensemble, where spectra are real and decoupled from the eigenvectors, with the result 
$H_{Z_tZ_t^{\dagger}}=z^2p^2-zp$, i.e.  identical to the Hamiltonian for the unitary diffusion \eqref{HamJW}, modulo factor $-1/2$.  
The corresponding HJ equations for both ensembles are related by replacing time $t$ in unitary diffusion by  $t \rightarrow -t/2$ for singular values evolution, pointing at some {\it a priori}  unexpected dualities between these two models. Such model  has also practical applications, in particular in the study of trainability of residual neural networks~\cite{RESNETS}.

\section{Asymptotics of Harish-Chandra-Itzykson-Zuber integral}
\label{sec:bridge}
Finally, presented formalism offers an appealing way to study the  asymptotics of the celebrated HCIZ~\cite{HC,IZ} integral:
\begin{align*}
I_{\text{HCIZ}} = \int dU e^{\frac{\beta}{2} N \text{Tr} U A U^\dagger B},
\end{align*}
for fixed matrices $A,B$ and parameter $\beta$ encoding whether integral is taken over unitary $\beta=2$ or orthogonal matrices $\beta = 1$. 

Following~\cite{MATYTSIN}, the main asymptotic contribution in the $N\to \infty$ limit reads
\begin{align*}
    I_{\text{HCIZ}} \sim e^{-\frac{\beta}{2} N^2 S},
\end{align*}
where the Euler-type hydrodynamic action reads
\begin{align}
\label{euleraction}
    S = \frac{1}{2} \int_0^1 dt \int dx \rho \left [ \mu^2 + \frac{\pi^2}{3} \rho^2 \right ],
\end{align}
with "fluid" density $\rho$ and momentum profile $\mu$. Action is evaluated on fluid trajectory such that  initial $\rho(x,t=0) = \rho_A(x)$ and final density $\rho(x,t=1) = \rho_B(x)$ are specified by matrices $A$ and $B$ respectively. Proper density is found by solving the following Euler equations:
\begin{align}
    \partial_t \mu + \mu \partial_x \mu & = \frac{\pi^2}{2} \partial_x (\rho^2), \label{euler1}\\
    \partial_t \rho + \partial_x ( \rho \mu) & = 0. \label{euler2}
\end{align}
One possible way pursued in \cite{MATYTSIN} is to compose a complex solution $h = \mu + i \pi \rho $ for which we recreate a well-known Burgers' equation:
\begin{align}
\label{hfunc}
\partial_t h + h\partial_z h = 0.    
\end{align}
Its solutions should then obey boundary conditions $\text{Im}\, h(t=0) = \pi \rho_A$ and $\text{Im}\, h(t=1) = \pi \rho_B$. Such an approach, while elegant, was of limited use to solve either special cases~\cite{MAJUMDAR} or as a method of indirect generation of solutions~\cite{MATYTSIN}. 

Our approach of finding proper solutions $h$ is slightly different and and has two stages -- first we look for a proper density $\rho$ based on the solution of bridge-type matrix dynamics and then match a velocity function $\mu$ such that the Euler equations \eqref{euler1} and \eqref{euler2} close.
\subsubsection{Brownian bridge matrix dynamics}
Finding a proper density consists of constructing matrix dynamics $Y_t$ starting at $Y_0=A$ and reaching matrix $Y_1 = B$. We first recall basic facts for one-dimensional stochastic process with such properties known as the \emph{bridge}.

We start from Brownian bridge dynamics for single variable \cite{SFPbridge} for which SFP equation starting at $x_0, t=0$ and ending in $x=x_f, t=t_f$ reads:
\begin{align*}
    \partial_t P(x,t) = D \partial_x \left [ \partial_x P -  2 \partial_x \ln Q(x,t) P \right ],
\end{align*}
where $Q$ is a solution to inverse SFP equation $\partial_t Q = - D \partial_{xx} P$ with $x=x_f$ for $t=t_f$ and reads $Q(x,t) = (4\pi D (t_f-t))^{-1/2} e^{-\frac{(x_f-x)^2}{4D(t_f - t)}}$. Therefore, SFP equation for the Brownian bridge reads
\begin{align*}
    \partial_t P = D \partial_{xx} P -  \partial_x \left ( \frac{x_f - x}{t_f - t} P \right ).
\end{align*}
Besides the usual diffusive term $\sim \partial_{xx} P$, there is a linear restoring force vanishing at $x=x_f$ and singular at $t = t_f$ ensuring that the trajectory ends up at the prescribed final point of the trajectory.

We move on to multi-dimensional generalization of the bridge process. We first decompose Hermitian matrix $Y_{kl} = x_{kl} + i y_{kl}$ for $k\neq l$, $Y_{kk} = x_{kk}$ and from above equation form a set of SFP formulas for each matrix element:
\begin{align*}
    \partial_t P(x_{ii},t) & = \frac{1}{2N} \partial_{x_{ii}}^2 P(x_{ii},t) - \partial_{x_{ii}} \left ( \frac{x_{ii}^f - x_{ii}}{t_f - t} P(x_{ii},t) \right ), \\
        \partial_t P(v_{ij},t) & = \frac{1}{4N} \partial_{v_{ij}}^2 P(v_{ij},t) - \partial_{v_{ij}} \left ( \frac{v_{ij}^f - v_{ij}}{t_f - t} P(v_{ij},t) \right ), \qquad i \neq j
\end{align*}
where $v = x,y$ and $v^f$ denote final matrix elements. Joint PDF $\mathcal{P}(Y,t) = \prod_i P(x_{ii},t) \prod_{i<j} P(x_{ij},t) P(y_{ij},t)$ satisfies a joint SFP equation $\partial_t \mathcal{P} = \mathcal{A} \mathcal{P}$ with
\begin{align*}
    \mathcal{A} & = \sum_{k=1}^N \left ( \frac{1}{2N} \partial_{x_{kk}}^2 P(x_{kk},t) - \partial_{x_{kk}} \frac{x_{kk}^f - x_{kk}}{t_f - t} \right ) + \nonumber\\
    & + \frac{1}{4N} \sum_{i<j} \left (  \partial_{x_{ij}}^2 + \partial_{y_{ij}}^2 \right ) + \\
    & - \sum_{i<j} \left ( \partial_{x_{ij}} \frac{x_{ij}^f - x_{ij}}{t_f - t} + \partial_{y_{ij}} \frac{y_{ij}^f - y_{ij}}{t_f - t} \right ).
\end{align*}

\subsubsection{Hamiltonian for the Brownian bridge matrix dynamics}
To find the Hamiltonian for the HJ equation of matrix bridge process, we first define a deformed  characteristic polynomial
\begin{align*}
    \hat{U}(z,\alpha,t) = \left < \det ( z - Y_t + \alpha B) \right >.
\end{align*}
Deformation consists of an arbitrary addition of auxiliary parameter $\alpha$ which has a role in what follows. Using a standard approach of \cite{tricks1,tricks2} we derive an exact dynamical equation for $\hat{U}$:
\begin{align*}
    \partial_t \hat{U} = \frac{1}{t_f - t} \left [ z \partial_z \hat{U} + (\alpha -1 )\partial_\alpha \hat{U} - N \hat{U} \right ] - \frac{1}{2N} \partial_{zz} \hat{U}.
\end{align*}
Parameter $\alpha$ is indispensable to close above equation. We transform it through half of the Cole-Hopf transform $\hat{\phi} = \frac{1}{N} \ln \hat{U}$ and take the large $N$ limit
\begin{align}
\label{bridgeeq2}
    \partial_t \hat{\phi} = - \frac{1}{2} (\partial_z \hat{\phi})^2 - \frac{1}{t_f - t} \left [ 1-z \partial_z \hat{\phi} - (\alpha -1)\partial_\alpha \hat{\phi}  \right ].
\end{align}
By the self-averaging property of the large $N$ limit, $\ln \hat{U} = \ln \left < \det (\cdots) \right > \sim \left < \ln \det (\cdots) \right > $, thus $\hat{\phi}$ becomes an effective potential $\hat{\phi}(z,\alpha,t) = \frac{1}{N} \left < \ln \det (z-Y_t + \alpha B)\right >$. Equation \eqref{bridgeeq2} is in the Hamilton-Jacobi form from which we read the Hamiltonian:
\begin{align}
\label{Hamsys}
    H_{\text{bridge}} = \frac{1}{2} p^2 + \frac{1}{t_f - t} \left [ 1 - zp - (\alpha-1)p_\alpha \right ],
\end{align}
where, besides the usual pair $z,p = \partial_z \hat{\phi}$, an auxiliary coordinate $\alpha$ and momentum $p_\alpha = \partial_\alpha \hat{\phi}$ is present. 

\begin{table*}[ht]
\caption{\label{tab1} Dictionary between optics/classical mechanics and random matrices}
\begin{ruledtabular}
\begin{tabular}{ccc}
optics & classical mechanics & (Hermitian/non-Hermitian) random matrices \\
\hline
real space & real space + time & complex/quaternionic space + time \\
Huygens principle & Hamilton-Jacobi equation & Hamilton-Jacobi equations \eqref{eq:HJHerm}/\eqref{eq:NHHJ} \\
ray & trajectory & characteristic curve \\
geodetic distance (eikonal) & action & electrostatic potential \eqref{potential} \\
slowness vector of the wavefront & momentum & Green's function \\
refractive index& Hamiltonian & integral of R-transform \eqref{main}

\end{tabular}
\end{ruledtabular}
\end{table*}

\textit{Solving HJ equation.}
Hamilton equations read
\begin{align*}
    \dot z & = p - \frac{z}{t_f-t}, \quad \dot \alpha = - \frac{\alpha-1}{t_f-t}, \\
    \dot p & = \frac{p}{t_f-t}, \quad \dot p_\alpha = \frac{p_\alpha}{t_f-t}.
\end{align*}
To continue, we first set $t_f - t = e^{-\tau}$ so that $(t_f - t) \frac{d}{dt} = \frac{d}{d\tau}$ and $t=0$ corresponds to $\tau = -\ln t_f$ while $t=t_f$ is transformed to $\tau \to \infty$. Hamilton equations are then
\begin{align*}
    \dot z & = e^{-\tau} p - z, \quad \dot \alpha = 1-\alpha, \\
    \dot p & = p, \quad \dot p_\alpha = p_\alpha.
\end{align*}
with the overdot denoting now $d/d\tau$. In the newly introduced time variable, the latter two equations for momenta are readily solved:
\begin{align*}
    p & = t_f e^\tau p_0 = \frac{t_f}{t_f-t} p_0, \\
    p_\alpha & = t_f e^\tau p_{\alpha,0}  = \frac{t_f}{t_f-t} p_{\alpha,0}.
\end{align*}
where $p_0 = p(\tau=-\ln t_f) = p(t=0), p_{\alpha,0} = p_\alpha(\tau=-\ln t_f) = p_\alpha(t = 0)$. We plug above solutions to the remaining Hamilton equations for $z,\alpha$:
\begin{align}
    \frac{dz}{t_f p_0 - z} = d\tau, \quad \frac{d\alpha}{1-\alpha} = d\tau,
\end{align}
and find the remaining solutions:
\begin{align*}
    z & = p_0(t_f - e^{-\tau}) + \frac{z_0}{t_f} e^{-\tau} = p_0 t + z_0 \frac{t_f-t}{t_f}, \\
    \alpha & =  1 - \frac{1}{t_f} e^{-\tau} + \frac{\alpha_0}{t_f} e^{-\tau} = \frac{t}{t_f} + \alpha_0 \frac{t_f-t}{t_f}.
\end{align*}
Next we introduce an initial condition $p_0(z_0,\alpha_0) = \left [ \partial_z \hat{\phi}\right ]_{\alpha=\alpha_0,z=z_0} = \frac{1}{N} \text{Tr} \frac{1}{z_0 - A + \alpha_0 B}$ which couples together $\alpha_0,p_0$ and $z_0$. We invert $p_0 = p (t_f-t)/t_f$ and calculate $\alpha_0 = 1 + (\alpha - 1)t_f/(t_f-t), z_0 = z t_f/(t_f-t) - pt$ so that the solution is given by
\begin{align*}
    p = t_f e^\tau p_0 \left (z_0 = \frac{z t_f}{t_f-t} - pt, \alpha_0 =  1 + \frac{(\alpha - 1)t_f}{t_f-t} \right ).
\end{align*}
or with explicitly plugged initial condition
\begin{align*}
    p = \frac{1}{N} \text{Tr} \frac{1}{z  - p \frac{t(t_f - t)}{t_f} - \frac{t_f - t}{t_f} A + \left ( \alpha - \frac{t}{t_f} \right )B}.
\end{align*}

In what follows we set $t_f=1, \alpha=0$ and reintroduce Green's function $G = p$. Define bridge function $G_{\text{Br}}(z,t) = \frac{1}{N} {\rm Tr} \frac{1}{z- (1-t)A - tB}$ and the resolvent is implicitly given by:
\begin{align}
\label{Gbridge}
G = G_{\text{Br}} \left (z - t(1-t)G,t \right).
\end{align}
By construction, the limits $t\to 0, 1$ recreate correct boundary Green's functions $G(z,t=0) = \frac{1}{N} \text{Tr} (z-A)^{-1}$ and $G(z,t=1) = \frac{1}{N} \text{Tr} (z-B)^{-1}$. At the same time, one can check explicitly that $G$ is itself not the sought solution to Burgers' equation \eqref{hfunc}. This discrepancy can be understood by decomposing $G$ into real and imaginary parts $G = i\pi \rho + \mathcal{H} \rho$ related by the Hilbert transform and ultimately dependent on single function $\rho$. On the other hand, the complex $h$ a priori consists of two independent (i.e. not related through any transform) functions $\rho,\mu$. We therefore assume that the densities are calculated correctly $\rho$ while the velocity profile needs further specification. Hence, in the second step we plug the density found from \eqref{Gbridge} into Euler equation \eqref{euler1} and calculate a matching velocity profile $\mu$. Once both $\rho$ and $\mu$ are identified, we evaluate the corresponding action \eqref{euleraction} and find the asymptotics of the HCIZ integral.

\textit{An example.} As a demonstration of the method we calculate the simplest example of vanishing matrices $A=B=0$. Boundary resolvent reads $G_{\text{Br}}(z_0) = 1/z_0$ and the density is simply a semicircle law $\rho_{\text{sem}}(x,t) = \frac{1}{2\pi t(1-t)} \sqrt{4t(1-t) - x^2}$ with appropriately rescaled size vanishing at both $t=0,1$. Plugging it into \eqref{euler1} results in:
\begin{align}
\label{eqeuler}
\partial_t \mu + \mu \partial_x \mu & = - \frac{x}{4t^2(1-t)^2},
\end{align}
since $\frac{\pi^2}{2} \partial_x (\rho_{\text{sem}}^2) = - \frac{x}{4t^2(1-t)^2}$. We solve \eqref{eqeuler} using method of characteristics
\begin{align*}
    \frac{d}{d\beta} \mu(\alpha,\beta) & =  - \frac{x}{4t^2(1-t)^2}, \\
    \frac{d}{d\beta} x(\alpha,\beta) & = \mu, \\
    \frac{d}{d\beta} t(\alpha,\beta) & = 1,
\end{align*}
with initial conditions $t(\alpha,0) = 1/2, x(\alpha,0) = \alpha$ and $v(\alpha,0) = 0$. We readily find $t=\beta +1/2$, plug into the remaining equations and combine them in matrix form
\begin{align*}
    \frac{d}{d\beta} \left ( \begin{matrix} \mu \\ x \end{matrix} \right ) = \left ( \begin{matrix} 0 & - \frac{1}{2(1-4\beta^2)^2} \\ 1 & 0 \end{matrix} \right ) \left ( \begin{matrix} \mu \\ x \end{matrix} \right ) .
\end{align*}
Solution is found by diagonalization:
\begin{align*}
 \mu(x,t) = \frac{2t-1}{2t(1-t)} x.   
\end{align*}
This example recreates the results of~\cite{MAJUMDAR}.

Although the above example only recreates the results of~\cite{MAJUMDAR}, approach by the HJ equation is general and does not require any guess-work. It provides a principled way of studying asymptotics of Berezin-Karpelevich integrals where similar hydrodynamic description was found \cite{GRELAFORRESTER} and non-Hermitian analogues of (generally unknown) HCIZ-type integrals.

\section{Summary}
We have proposed to apply the Hamilton-Jacobi dualism between the Lagrange-Euler description (based on trajectories) and the Hamilton  description (based on wavefronts)  in the context of large $N$ dynamical random matrix models. 
As shown in Tab. \ref{tab1}, we have successfully transferred the optical analogy between the Fermat principle  and the Huygens principle to the realm of large random matrices. 
Such a scheme, is to the best of our knowledge, novel and offers an inspiring  perspective for merging
several physical concepts from  classical mechanics,  optics,  hydrodynamics, statistical physics  and even  quantum mehcanics with advanced mathematical methods of random matrix theory. 

As a result of this approach, we derive several formulas. Firstly, we present HJ equations \eqref{eq:HJHerm} and \eqref{eq:NHHJ} for general (i.e. Hermitian and non-Hermitian as well as non-Gaussian) dynamics. Secondly, we deliver concrete Hamiltonians for Ornstein-Uhlenbeck \eqref{HOU}, elliptic \eqref{Helliptic}, bi-unitary \eqref{Hbiunit} matrix processes and several other as summarized in Tabs. \ref{tab1a},\ref{tab2a}. Lastly, we enlarge the matrix HJ formalism to a matrix bridge scenario \eqref{Hamsys} enabling recreation of the asymptotics of Harish-Chandra-Itzykson-Zuber integral.

We believe that, since the most interesting phenomena in random matrix models  (e.g. new classes of universalities) occur mostly at places where wavefronts change their behaviour (boundaries of the spectral support, corresponding to gradient catastrophes), the formalism  which focuses on  such objects is indeed promising. 

In particular, in future work we plan to apply quantization of HJ equation \cite{HJQUANT} as a fresh approach to universality achieved directly from large $N$ asymptotics. Another advantage of this formulation is visible at the level of non-normal models, where the proper identification of canonical "coordinates" and "momenta" leads to complete treatment of the evolution of both eigenvalues and the eigenvectors. The resulting HJ equations represent a  dimensional reduction of large $N$ problems. 
The proposed formalism allows also for rephrasing several  open questions, like the issue of large deviations in non-normal matrix models, which we plan to expose in the sequel to this  work.

\begin{acknowledgments}
\textit{Acknowledgments.} The authors are indebted to Todd Kemp and Brian Hall (whose research have  triggered our interest in the physical foundations of the HJ dynamics) for discussions and correspondence. The authors also thank Romuald Janik for helfpul remarks. 
The research was supported  by the TEAMNET POIR.04.04.00-00-14DE/18-00 grant of the Foundation for Polish Science. WT was also supported by ETIUDA scholarship UMO2018/28/T/ST1/00470 from National Science Center.
\end{acknowledgments}

\appendix
\section{Fig. 1b - Characteristics and wavefronts for GUE dynamics}
\label{guedyn}
We find the potential function $\phi$ by solving HJ equation \eqref{eq:HJHerm}:
\begin{align}
    \partial_t \phi + H(p=\partial_z \phi) = 0
\end{align}
for the simplest case of GUE dynamics $H(p) = p^2/2$. Next, we use obtained formulas to plot wavefronts alongside characteristics as their natural counterparts and comment on how Fig. \ref{fig1}b in the main text was obtained. 

The Hamiltonian does not depend explicitly dependent on time so that the solution is of the form
\begin{align}
    \phi(z,\alpha,c,t) = W(z,\alpha,c) - \alpha t,
\end{align}
where $\alpha$ is the first conserved quantity and $c$ is an overall additive constant. HJ equation \eqref{eq:HJHerm} reads
\begin{align}
    \alpha = \frac{1}{2} \left ( \partial_z W \right )^2
\end{align}
so that $W = \sqrt{2\alpha} z + c$ and the full solution reads:
\begin{align}
    \phi(z,\alpha,t,c) = \sqrt{2\alpha} z(t) - \alpha t + c.
    \label{solution00}
\end{align}
Canonical transformations give the transformed constant position $\beta$ and "old" momentum $p$:
\begin{align*}
    \beta & = \partial_\alpha \phi = \frac{z}{\sqrt{2\alpha}} - t,\nonumber \\
    p & = \partial_z \phi = \sqrt{2\alpha}.
    \label{canon}
\end{align*}
Position $z$ and momenta $p$ are given in terms of constants of motion $\alpha,\beta$ as:
\begin{align*}
    z(t) & = \sqrt{2\alpha} (\beta + t) \\
    p(t) & = \sqrt{2\alpha}
\end{align*}
Constants of motion are reformulated in terms of initial position $z(0) = z_0 = \beta \sqrt{2\alpha}$ and momentum $p(0)= p_0 = \sqrt{2\alpha}$ or $\alpha = p_0^2/2, \beta = z_0/p_0$. We plug it back to the solution \eqref{solution00}:
\begin{align*}
    \phi = p_0 z_0 + c + \frac{p_0^2}{2} t.
\end{align*}
In terms of initial conditions, position and momentum describe a one-dimensional free particle:
\begin{align*}
    z(t) & = z_0 + p_0 t, \\
    p(t) & = p_0.
\end{align*}
In our case, initial positions and momenta $z_0, p_0$ are related by the initial condition $\phi(z_0,t=0) = f(z_0)$ and $p_0 = \partial_z \phi(t=0)_{|z=z_0} = f'(z_0)$. These relations between $z_0, p_0$ are:
\begin{align*}
    p_0 z_0  +c & = f(z_0), \\
    p_0 & = f'(z_0),
\end{align*}
which results in the final form of the potential function:
\begin{align*}
    \phi(z(t),t) = f(z_0) + \frac{(z(t)-z_0)^2}{2t},
\end{align*}
where $z(t) = z_0 + f'(z_0) t$. It gives only the holomorphic part of the function while the total potential reads:
\begin{align}
    \Phi(z(t),t) = \phi + \bar{\phi} = 2 \text{Re} f(z_0) + \frac{1}{t} \text{Re} (z(t)-z_0)^2.
\end{align}
\paragraph{Obtaining Fig. 1b.} Figure in the main text is found by plotting wavefronts as equipotential surfaces $\Phi(z(t),t) = \text{const}$ while complex characteristics form a family of curves given by $z(t) = z_0 + f'(z_0) t$ where $z_0$ is the parameter labeling the curves. Both were found for a special case $f(z) = \log z$.

\begin{figure}
    \centering
    \includegraphics[scale=0.28]{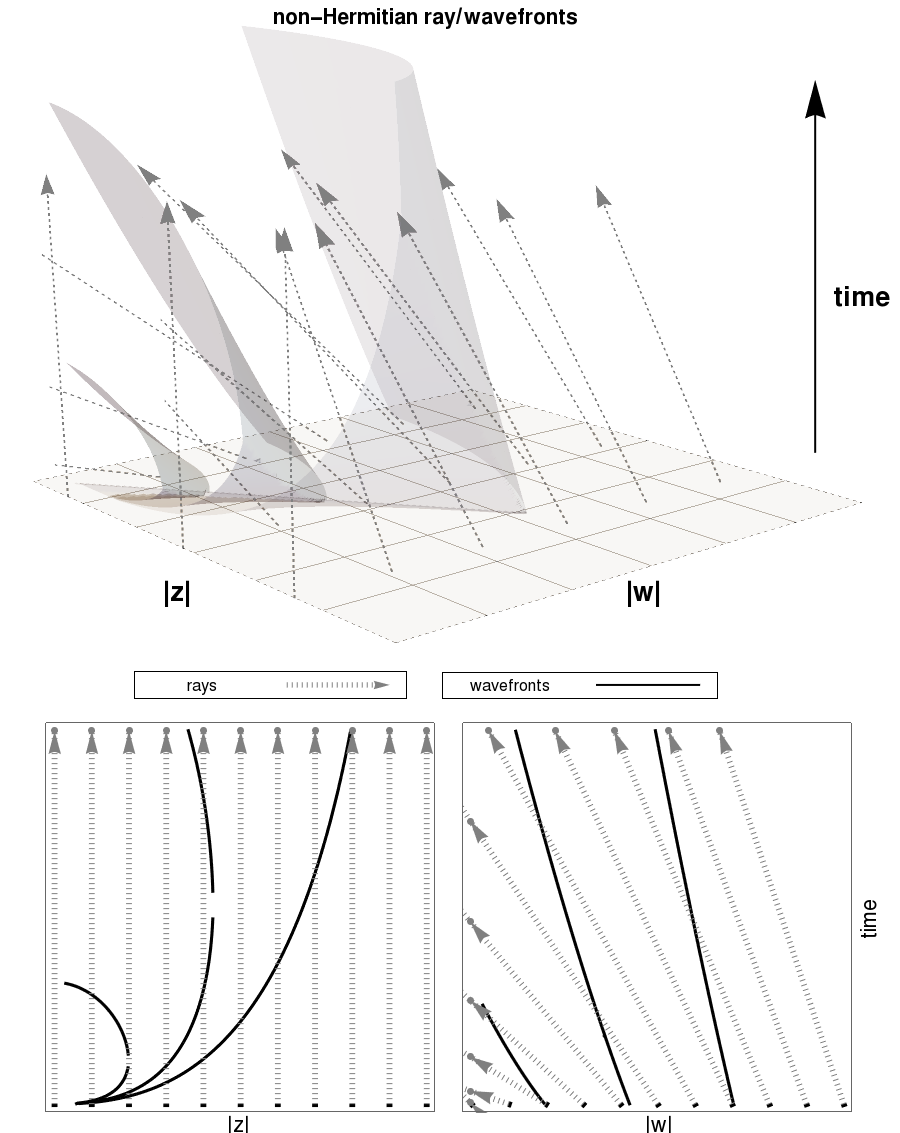}
    \caption{Construction of duality between wavefronts (black solid lines) and rays (dotted gray arrows) in non-Hermitian dynamical random matrices. Propagation takes place in restricted quaternionic space spanned by modules $|z|$ and $|w| = r$. Wavefronts are defined by condition $\Phi(r(t),t) = \text{const}$ with potential given by Eq. \eqref{fin} and characteristics $z(t) = z_0, r(t) = r_0 - r_0/(|z_0|^2+r_0^2) t$ by Eq. \eqref{req}. $z$ variable is trivially added despite absence of interesting dynamics. Similarly to Fig. 1 in the main text, propagation is likewise anisotropic. Presented evolution happens for the Ginibre dynamics.}
    \label{fig2}
\end{figure}

\section{Characteristics and wavefronts for Ginibre dynamics}
In this section we solve HJ equation:
\begin{align}
    \partial_t \Phi + H \left ( \P = (\dQ \Phi)^T, \Q \right ) = 0,
\end{align}
for Ginibre dynamics with $H = - |p_w|^2$. We essentially follow the same steps as in App. \ref{guedyn} on GUE dynamics. As a result, in Fig. \ref{fig2} we present both wavefronts and characteristics in a restricted quaternionic space spanned by moduli $|z|,|w|=r$ and time $t$.

In polar coordinates $w = re^{i\alpha}$, Ginibre Hamiltonian reads $H(r,p_r,t) = - \frac{1}{4} p_r^2$. Notice that despite the variable $z$ missing in the Hamiltonian, the dynamics still takes place in both $z$ and $w$.

Since the Hamiltonian does not depend on time, we start off by setting:
\begin{align*}
    \Phi(r,\alpha,c,t) = W(r,\alpha,c) + \alpha t,
\end{align*}
where $\alpha$ is the first conserved quantity and $c$ is an overall additive constant. We plug in the ansatz to HJ equation and find
\begin{align*}
    \alpha = \frac{1}{4} \left ( \partial_r W \right )^2
\end{align*}
so that solution reads $W = \sqrt{4\alpha} r + c$. Full potential $\Phi$ reads:
\begin{align}
    \Phi(r,\alpha,t) = \sqrt{4\alpha} r(t) + \alpha t + c
    \label{solution}
\end{align}
where we write down explicitly $r$'s relation on time. Canonical transformations give the transformed constant position $\beta$ and "old" radial momentum $p_r$:
\begin{align}
    \beta & = \partial_\alpha \Phi = \frac{r}{\sqrt{\alpha}} + t,\nonumber \\
    p_r & = \partial_r \Phi = \sqrt{4\alpha}.
    \label{canon}
\end{align}
What results are equations of motion:
\begin{align*}
    r(t) & = \sqrt{\alpha} (\beta - t), \\
    p_r(t) & = 2 \sqrt{\alpha},
\end{align*}
given in terms of constants of motion $\alpha,\beta$ (or new position/momenta). Next, we evaluate these constants in terms of interpretable quantities like initial position and momentum $r(t=0)=r_0,p_r(t=0)=p_{r,0}$. We set $t=0$ in both equations and obtain
\begin{align*}
    r_0 = \sqrt{\alpha} \beta, \\
    p_{r,0} = 2\sqrt{\alpha},
\end{align*}
and solve for $\alpha,\beta$ to obtain $\alpha = \frac{1}{4} p_{r,0}^2, \beta = \frac{2r_0}{p_{r,0}}$. We plug those back into \eqref{canon} and solve them to obtain:
\begin{align*}
    r(t) & = r_0 - \frac{p_{r,0}}{2} t \\
    p_r(t) & = p_{r,0}.
\end{align*}
In this parametrization, the Hamilton principal function (or potential) reads
\begin{align}
\label{F2p0q0}
\Phi = p_{r,0} r + \frac{1}{4} p_{r,0}^2 t + c.
\end{align}
Again, problem we aim to solve couples inital momentum and position through initial value of the potential $t=0$ $\Phi(r_0,\alpha,t=0) = F(r_0)$ and its derivative $\partial_r \Phi(r_0,\alpha,t=0) = F'(r_0)$. These two conditions translate to relation between initial $r_0, p_0$ and constant $c$:
\begin{align*}
    \Phi(t=0) = F(r_0) & \quad \to \quad r_0 p_{r,0} + c = F(r_0), \\
    \partial_r \Phi(t=0) = F'(r_0) & \quad \to \quad p_{r,0} = F'(r_0).
\end{align*}
This in turn renders the underlying $r,p$ dynamics dependent only on $r_0$:
\begin{align}
    r(t) & = r_0 - \frac{F'(r_0)}{2} t, \label{req}\\
    p(t) & = F'(r_0),
\end{align}
and sets the constant $c = F(r_0) - F'(r_0) r_0$. Lastly, we plug all newfound quantities into $\Phi$ given by \eqref{F2p0q0}:
\begin{align}
\label{fin}
    \Phi(r(t),t) = F(r_0) - \frac{(r(t) - r_0)^2}{t},
\end{align}
where $r_0$ is expressed through $r(t) = r_0 - \frac{F'(r_0)}{2} t$. 

We consider the simplest initial value $F(r_0) = \log (|z|^2 + r_0^2)$. Both wavefronts and characteristics are plotted in Fig. \ref{fig2}.

 \vfil

\end{document}